\documentclass[12pt]{article}
\newcommand{\Comment}[1]{{}}
\usepackage[textwidth = 450 pt, textheight = 630 pt]{geometry}
\usepackage{amssymb,euscript,amsmath,amsfonts}
\usepackage{tikz,tikz-cd,url}
\usetikzlibrary{decorations.pathreplacing,calligraphy}
\usepackage{xcolor}
\usepackage{graphicx}
\usepackage{color}

\definecolor{MyDarkBlue}{rgb}{0.15,0.15,0.45}
\usepackage[linktocpage=true]{hyperref}
\hypersetup{
colorlinks=true,
citecolor=MyDarkBlue,
linkcolor=MyDarkBlue,
urlcolor=MyDarkBlue,
pdfauthor={Neil Lambert },
pdftitle={ },
pdfsubject={hep-th}
}

\usepackage[numbers,sort&compress]{natbib}
\usepackage{hypernat}
\usepackage{accents}
\usepackage{bm}

\begin{document}

\renewcommand{\thefootnote}{\fnsymbol{footnote}}

   \vspace{1.8truecm}

 \centerline{\LARGE \bf {\sc Six-Dimensional Correlators From a } }
 \vskip12pt
 
 \centerline{\LARGE \bf {\sc Five-Dimensional Operator Product Expansion}}
 
 


\vspace{1cm}

\vspace{1.0truecm}

 
\thispagestyle{empty}

\vspace{0cm}
  \centerline{
   {\large {\bf  {\sc N.~Lambert,${}^{\,a}$}}\footnote{E-mail address: \href{neil.lambert@kcl.ac.uk}{\tt neil.lambert@kcl.ac.uk}}     \,{\sc A.~Lipstein,$^{\,b}$}\footnote{E-mail address: \href{mailto:arthur.lipstein@durham.ac.uk}{\tt arthur.lipstein@durham.ac.uk}}   and {\sc R.~Mouland${}^{\,c}$}\footnote{E-mail address: \href{rishi.mouland@kcl.ac.uk}{\tt r.mouland@damtp.cam.ac.uk}}}   }  
     
\vspace{1cm}
\centerline{${}^a${\it Department of Mathematics}}
\centerline{{\it King's College London }} 
\centerline{{\it London, WC2R 2LS, UK}} 
  
\vspace{1cm}
\centerline{${}^b${\it Department of Mathematical Sciences}}
\centerline{{\it Durham University}} 
\centerline{{\it  Durham, DH1 3LE, UK}} 

\vspace{1cm}
\centerline{${}^c${\it Department of Applied Mathematics and Theoretical Physics}}
\centerline{{\it University of Cambridge }} 
\centerline{{\it Cambridge, CB3 0WA, UK}}
  
\vspace{12pt}

\centerline{\sc Abstract}
\vspace{0.4truecm}
\begin{center}
\begin{minipage}[c]{360pt}{
    \noindent In this letter we discuss the operator product expansion of scalar operators in five-dimensional  field theories with an $SU(1,3)\times U(1)$ spacetime symmetry. Such theories arise by a novel conformal null reduction  of six-dimensional Lorentzian conformal field theories. Unlike Lorentzian conformal field theories, three-point functions of generic operators in such theories are not completely fixed by $SU(1,3)\times U(1)$ symmetry. However, we show that in a special case the functional form of the OPE coefficients can be fully determined, and we use them to fix the form of the three-point function. The result is shown to agree with correlation functions obtained by reduction of six-dimensional conformal field theories.}
 
\end{minipage}
\end{center}

\newpage 
 \section{Introduction}\label{sect: Intro}
 
Higher-dimensional conformal field theories (CFTs) are intrinsically strongly coupled and as such do not admit traditional Lagrangian descriptions. Therefore alternative techniques must be used to define and study them. These  include holography \cite{Bastianelli:1999ab,Bastianelli:1999vm,Eden:2001wg,Arutyunov:2002ff}, bootstrap methods \cite{Heslop:2004du,Rastelli:2017ymc,Heslop:2017sco,Abl:2019jhh,Alday:2020tgi}, chiral algebras \cite{Beem:2014kka,Chester:2018dga}, deconstruction \cite{Arkani-Hamed:2001wsh,Lambert:2012qy} or discrete light cone quantization \cite{Aharony:1997an,Hellerman:1997yu}. However to date there is no systematic approach to these theories. 
 
An alternative approach arises by considering a certain conformal mapping of six-dimensional (6D) Minkowski space that brings one of the two light cone coordinates, say $x^+$, into a finite range: $x^+\in (-\pi R,+\pi R)$ \cite{Lambert:2019jwi}, where $R$ is a constant with dimensions of length that is introduced by the coordinate transformation along with an anti-symmetric and self-dual tensor $\Omega_{ij}$ with $\Omega_{ij}\Omega_{jk}=-R^{-2}\delta_{ik}$. This can be done in such a way that $\partial_+$ remains a Killing direction. In this way we can perform a Kaluza-Klein reduction along $x^+$ while still studying the non-compact six-dimensional theory. The resulting  five-dimensional  (5D) theory has an  $SU(1,3)\times U(1)$ spacetime symmetry  group that commutes with  the momentum generator $P_+$ (which generates the $U(1)$ factor) \cite{Lambert:2019fne}. Due to the novel coordinate transformation used in the null compactification one finds that the $SU(1,3)\times U(1)$ symmetry is an $\Omega$-deformed version of the more familiar $z=2$ Schr\"odinger group of conventional non-relativistic conformal field theories \cite{Nishida:2007pj}. In particular the generators of $SU(1,3)$ are   $M_{i+},K_+,P_-,P_i, B, C^I,T$, which are linearly related  to the familiar generators $\hat{M}_{\mu\nu},\hat{K}_\mu,\hat{P}_\mu,\hat{D}$ of $SO(2,6)$. Here, $P_-$ is the Hamiltonian, and $P_i$ generates spatial translations. $B$ is the rotation generated by $\Omega_{ij}$, while $C^I$ are the remaining spatial rotations that commute with $\Omega_{ij}$. Finally, $T$ is Lifshitz scaling symmetry,   $M_{i+}$ is an $\Omega$-deformed Galilean boost, and $K_+$ is a special conformal generator.

Unlike correlation functions with Schr\"odinger symmetry, those constrained by $SU(1,3)\times U(1)$ have power-law suppression in space and time. Indeed, by assembling five-dimensional operators into Fourier series along the compact null direction, one can recover the correlation functions of six-dimensional CFTs in Minkowski space without having to take the limit $R\to\infty$ \cite{Lambert:2020zdc}. A key benefit of the theories with $SU(1,3)\times U(1)$ symmetry is that they admit Lagrangian descriptions, including theories with a large number of supersymmetries as appropriate for reduction of six-dimensional CFTs with $(2,0) $ or $(1,0)$ supersymmetry \cite{Lambert:2019jwi,Lambert:2020jjm}. These are rather non-standard gauge theories and the role of the $P_+$ eigenvalue is given by the instanton number. They admit a  Lifshitz scaling property and as such might be embedded into UV complete theories  \cite{Lambert:2022ztz}. Thus there is some hope that these Lagrangians give well-defined path integrals and can be used to compute correlation functions in six-dimensional CFTs and also elucidate the relation of abstract CFT to gauge field theories.

One of the most important concepts in conformal field theory is the operator product expansion (OPE), which expresses the product of two local operators located at different points as an infinite series of local operators located at one of the two points. The existence of such an expansion is a basic assumption of Lorentzian CFTs like the 6D $(2,0)$ theory, and we will assume that the same holds for local scalar operators in the 5D theories which arise from them via the reduction described above. Note that this may be a nontrivial assumption since such operators correspond to Fourier modes along the null direction and are therefore non-local from a 6D perspective. Using the 5D OPE, we then show that if one of the operators in a three-point function satisfies the constraint $p_+ = \pm \Delta/2R$, where $p_+$ is the $P_+$ eigenvalue and $\Delta$ is the scaling dimension, then the three-point function can be fully determined. Moreover, we show that this agrees with the result obtained from Fourier decomposing a six-dimensional Lorentzian CFT correlator along the compact null direction. Note that for generic operators, the 5D three-point functions are not completely fixed by $SU(1,3)\times U(1)$ \cite{Lambert:2020zdc}, so it appears that the 5D OPE can be used to deduce 6D correlators at least for certain operators.

Before proceeding, let us briefly point out an alternative interpretation of the correlation functions studied in this letter. Most generally, they can be understood as the correlation functions $\langle L_1 L_2 \dots L_n \rangle $ of a collection of line operators $L_i$ in a six-dimensional Lorentzian relativistic CFT, each extended along an integral curve of the same null conformal Killing vector field. This vector field is fixed by the parameter $R$, and in particular as we take $R\to \infty$, the lines follow parallel null geodesics. We expect our results to generalise to $SU(1,n)$ theories, corresponding to correlation functions of null line operators in $2n$-dimensional relativistic conformal field theory.
 
 
This letter is organised as follows. In section \ref{review}, we review some basic properties of the 5D theories and their symmetries. In section \ref{sect: OPE}, we consider the OPE of scalar operators in the 5D theories and show that $SU(1,3)\times U(1)$ symmetry can be used to fix the OPE coefficients for all the descendants of a primary operator appearing the expansion in terms of the coefficient of that operator, {\it i.e.} the leading OPE coefficient. In general, the leading OPE coefficient is an unfixed function, however we find that when $p_+ = \pm \Delta/2R$ all the OPE coefficients can be determined. In section \ref{3pt} we show that scalar three-point functions are determined by the leading coefficient in the OPE of two of the operators. Hence, we can fix the form of a three-point function if one of the operators in a three-point correlator satisfies $p = \pm \Delta/2R$, and we show that this form arises from Fourier expanding 6D Lorentzian correlators. In section \ref{sect: Conclusions} we give a short conclusion.

 
\section{Review} \label{review}

The five-dimensional theories we consider arise from placing the 6D Lorentzian CFTs on a manifold with metric
\begin{align}
ds^2 = -2dx^+ \!\left(dx^- - \frac12 \Omega_{ij}x^idx^j\right)  + dx^idx^i,
\label{metric}
\end{align}
where $i \in \left\{ 1,2,3,4\right\} $, $-\pi R \leq x^+ \leq \pi R$, and $\Omega$ is an anti-self-dual 2-form satisfying $\Omega_{ik} \Omega_{jk}=R^{-2}\delta_{ij}$. This metric  is conformally flat and can be obtained from a standard 6D Minkowski metric $ds^2 =d\hat{x}^{\mu} d\hat{x}_{\mu}$ via a change of variables and Weyl transformation \cite{Lambert:2020zdc}. Reducing along $x^+$ then gives rise to a five-dimensional theory that admits a Lagrangian description which can in principle be used to compute 6D observables non-perturbatively via path integrals \cite{Lambert:2021fsl}.  Although we will not explicitly need the Lagrangians, for the interested reader we display the bosonic part below \cite{Lambert:2019jwi}:
\begin{align}
\begin{split}
    \mathcal{L}_{bos} \propto \text{Tr} \Big\{ \frac{1}{2} F_{i -} F^{i}{}_{-} + \frac{1}{2}  G_{ij} \mathcal{F}^{ij} - \frac{1}{2} \nabla_{i} X^I \nabla^i X^I  \Big\},
    \end{split}
\label{lag}
\end{align}
where $\nabla_i = D_i - \frac{1}{2} \Omega_{i j} x^{j} D_-$, with $D_-$ and $D_i$ being standard covariant derivatives for the gauge fields $A_-$ and $A_i$, $G_{ij}$ is a self-dual Lagrange multiplier, $\mathcal{F}_{ij}$ is a  field strength constructed from a linear combination of the the field strengths $F_{ij}$ and $F_{-i}$, and $X^I$ are scalars. Here, $I$  is an R-symmetry index. 

The Lagrangian enjoys an $SU(1,3)\times U(1)$ spacetime symmetry, which is the isometry group of the metric in \eqref{metric} after reducing along the $x^+$ direction. From the 6D perspective the $U(1)$ is generated by $P_+=\partial_+$ while in the 5D Lagrangian theory it corresponds to the instanton number \cite{Lambert:2021fsl}. Among the 15 generators of $SU(1,3)$, the ones that we will primarily make use of in this paper are
\begin{align}
\left( M_{i+} \right)_\partial 	\ &= \ \left( \tfrac{1}{2}\Omega_{ij} x^- x^j - \tfrac{1}{8}R^{-2} |x|^2x^i \right)\partial_- + x^- \partial_i  +( \tfrac{1}{2}\Omega_{ik}x^k x^j + \tfrac{1}{2}\Omega_{jk}x^k x^i - \tfrac{1}{4}\Omega_{ij}|x|^2 )\partial_j	\, , \nonumber\\
	\left( K_{+} \right)_\partial 	 \ &= \ ( 2 ( x^- )^2 - \tfrac{1}{8} R^{-2} |x|^4 )\partial_- +( \tfrac{1}{2} \Omega_{ij} x^j |x|^2 + 2 x^- x^i )\partial_i	\ ,\nonumber\\
\left( T \right)_\partial \ & =\  2x^-\partial_- + x^i\partial_i		\ ,
\end{align}
where $|x|^2 = x^ix^i$. In the limit $R\rightarrow \infty$ these generators reduce to Galilean boosts, special Schr\"odinger transformations, and a Lifshitz scaling, respectively. For more details of the symmetry algebra, see \cite{Lambert:2019fne,Lambert:2021fsl}.

Primary operators are labelled by their Liftshitz dimension $\Delta$, and $P_+$ eigenvalue $p_+$ (in what follows we drop the subscript $+$ to clear up the notation), as well as their irreducible representation $(R_\mathcal{O}[B],R_\mathcal{O}[C^I])$ under the $U(1)\times SU(2)$ rotation subgroup. They are defined by the following transformation properties:
 \begin{align}
  	[\mathcal{O}(x),M_{i+}]	\ &= \	\left( M_{i+} \right)_\partial \mathcal{O}(x) + \left(\tfrac{1}{2}\Delta \Omega_{ij} x^j - ip x^i	 +\tfrac{2}{R}x^i R_\mathcal{O}[B] - \Omega_{ik} \eta^I_{jk} x^j R_\mathcal{O}[C^I]\right) \mathcal{O}(x)			\, \nonumber\\
  	[\mathcal{O}(x),K_{+}]			\ &=	 \ \left( K_+ \right)_\partial \mathcal{O}(x)  + \left(2\Delta\, x^- - ip |x|^2 +\tfrac{2}{R}|x|^2 R_\mathcal{O}[B] - x^i x^j \Omega_{ik}\eta^I_{jk} R_\mathcal{O}[C^I]   \right) \mathcal{O}(x) \, 	\nonumber\\
  	[\mathcal{O}(x),T]	\ &= \	\left( T \right)_\partial \mathcal{O}(x) + \Delta \mathcal{O}(x)			\, .
\label{primary1}
 \end{align}
We will only look at scalar operators so we can drop the $R_\mathcal{O}[C^I]$ and $R_\mathcal{O}[B]$ terms. Such operators then satisfy the conditions   
\begin{align} \label{primary2}
  	[\mathcal{O}(0),M_{i+}]	\ &= 0	\, \nonumber\\
  	[\mathcal{O}(0),K_{+}]			\ &= 0\\ 
  	[\mathcal{O}(0),T] \ & =  \Delta\mathcal{O}(0)\nonumber\ ,
  	\end{align}
Note that conservation of $P_+$ implies that the sum over all the $U(1)$ charges of the operators in a given correlator should vanish, {\it i.e.} $\sum_K p_K=0$. Although we will not need to be explicit about the scalar operators which appear in correlation functions, to get an idea of the kind of operators we have in mind, recall that in the 6D $(2,0)$ theory one can construct protected operators by taking the trace of a product of scalar fields $X^I$ which are symmetrised and traceless in the R-symmetry indices. Upon reduction to 5D, we once again get a trace of a product of $X^I$ fields but it will be dressed with an instanton operator which encodes the mode number along the compact null direction \cite{Lambert:2021fsl}. 

As shown in \cite{Lambert:2020zdc}, the two-point scalar correlators of the theory are completely fixed by the $SU(1,3)\times U(1)$ symmetry and take the form
\begin{align}\label{2pts}
\langle \mathcal{O}_I(x_1)\mathcal{O}_J(x_2)\rangle	=c_{IJ} \delta_{p_I,-p_J}\delta_{\Delta_I,\Delta_J}\left(\frac{1}{z_{12}\bar z_{12}}\right)^{\frac{\Delta_I}{2}}	\left(\frac{z_{12}}{\bar z_{12}}\right)^{Rp_I}\ ,
\end{align}
where for any two points
\begin{align}
z_{ab} = x_a^--x_b^- +\frac12\Omega_{ij}x^i_ax^j_b +\frac{i}{4R}(x_a^i-x^i_a)	(x^i_a-x^i_b)	\ ,
\end{align}
which satisfies $\bar z_{ab} = -z_{ba}$.  
Moreover, three-point scalar correlators are constrained by the $SU(1,3)$ Ward identities to take the form
\begin{align}\label{3gen2}
\langle \mathcal{O}_I(x_1)\mathcal{O}_J(x_2)\mathcal{O}_K(x_3)\rangle 
&= 	\left(\frac{1}{z_{12}\bar z_{12}}\right)^{\frac{\Delta_I+\Delta_J-\Delta_K}{4}}	\left(\frac{1}{z_{23}\bar z_{23}}\right)^{\frac{-\Delta_I+\Delta_J+\Delta_K}{4}}	\left(\frac{1}{z_{13}\bar z_{13}}\right)^{\frac{\Delta_I-\Delta_J+\Delta_K}{4}}\nonumber\\ &\qquad \times
\left(\frac{z_{12}}{\bar z_{12}}\right)^{\tfrac{R}{3}(p_I-p_J)}\left(\frac{z_{23}}{\bar z_{23}}\right)^{\tfrac{R}{3}(p_J-p_K)}\left(\frac{z_{13}}{\bar z_{13}}\right)^{\tfrac{R}{3}(p_I-p_K)}	\nonumber\\
& \qquad\times H_{IJK}\left(\frac{z_{21}z_{23}z_{31}}{\bar z_{12}\bar z_{23}\bar z_{31}}\right)\delta_{-p_K,p_I+p_J}\ ,
\end{align}
for an unknown function $H_{IJK}$ of the single variable $\zeta=z_{12}z_{23}z_{31}/{\bar z_{12}\bar z_{23}\bar z_{31}}$.

It was shown in \cite{Lambert:2020zdc} that the above forms for the 2-point (\ref{2pts}) and 3-point functions (\ref{3gen2}) are consistent with the correlation functions of local operators in the initial $SO(2,6)$-invariant 6D theory, with the $p_I$ corresponding to momentum along a conformally compactified sixth direction, as they must be. That is, we can dimensionally reduce the well-known form of 2- and 3-point functions to obtain 5D correlators of the above form. In particular, the functions $H_{IJK}$ are entirely fixed, up to an overall factor of the relevant 6D OPE coefficient, as we review in detail in Section \ref{subsec: dim red}. Conversely, it was shown that we can perform a Fourier resummation of such 5D correlators---taking care of certain ordering ambiguities---to recover their 6D counterparts. This is the essential mechanism by which one might study 6D correlation functions from a 5D perspective.

\section{5D Operator Product Expansion}\label{sect: OPE}
 
In general, the operator product expansion of two primary operators can be written as a sum over primaries and descendants, which are obtained by acting with derivatives on the primaries. The coefficients in this sum are known as OPE coefficients. Our goal in this section will be to show that $SU(1,3)$ symmetry can be used to fix all the descendent OPE coefficients in terms of the primary OPE coefficients. This is a standard result in relativistic CFT, and was extended to theories with Schr\"odinger symmetry in \cite{Golkar:2014mwa,Goldberger:2014hca}. Following those references, we consider an OPE of scalar primary operators of the form
 \begin{align}\label{ope}
 \mathcal{O}_I(x)\mathcal{O}_J(0) &= \sum_{K,{\vec n},m} C_{IJ   }^{K,{\vec n}, m}(x)\partial_{\vec n}\partial_-^m\mathcal{O}_K(0)\nonumber\\
 &= \sum_K\bigg( C^{K, \vec 0, 0}_{IJ}(x)\mathcal{O}_K(0) + C^{K i, 0}_{IJ}(x)\partial_i \mathcal{O}_K(0)\nonumber\\	&\qquad \qquad \qquad +  C^{K ij,0}_{IJ}(x_I)\partial_i\partial_j\mathcal{O}_I(0) + C^{K \vec, 0, 1}_{IJ}(x)\partial_- \mathcal{O}(0) +\ldots\Big)\ ,
 \end{align}
 where the sum is over primary operators $\mathcal{O}_K$ as well as their descendants $\partial_{\vec n}\partial_-^m\mathcal{O}_K $.  Here $\vec n$ is short hand for a string of partial spatial derivatives with length $|\vec n|$ and $m=0,1,2,...$ counts the number of   $\partial_-$ derivatives. In the second line we have explicitly written out the first four terms for illustration.

The idea is to commute $T$, $M_{i+}$ and $K_+$ given in section \ref{review} with the left and right hand side of (\ref{ope}) and then evaluate the left hand side using (\ref{ope}). This leads to a set of differential equations that relate the various coefficients  $C^{K,i,0}_{IJ},C^{K,kl,0}_{IJ}$ {\it etc.} to $C^{K, \vec 0,0}_{IJ}$. 

For example we can consider $T$. Using \eqref{primary1} and \eqref{primary2} on the left hand side of \eqref{ope} we find
\begin{align}
[T,\mathcal{O}_I(x)\mathcal{O}_J(0) ]	 & = [T,\mathcal{O}_I(x)]\mathcal{O}_J(0) 	+\mathcal{O}_I(x)[T,\mathcal{O}_J(0) ]	\nonumber\\
& = -(\left( T \right)_\partial \mathcal{O}_I(x) +\Delta_I \mathcal{O}_I(x))\mathcal{O}_J(0) - \mathcal{O}_I(x)(\Delta_J\mathcal{O}_J(0))\nonumber\\
& = -\left( T \right)_\partial \mathcal{O}_I(x)\mathcal{O}_J(0) - (\Delta_I+\Delta_J)\mathcal{O}_I(x)\mathcal{O}_J(0)\nonumber \\
& =  -\sum_{K,{\vec n},m} \left( T \right)_\partial C^{K ,{\vec n},m}_{IJ}(x)\partial_{\vec n}\partial_-^m\mathcal{O}_K(0)- (\Delta_I+\Delta_J)\sum_{K,{\vec n},m} C^{K {\vec n},m}_{IJ}(x)\partial_{\vec n}\partial_-^m\mathcal{O}_K(0)\ .\end{align}
Note that $ \left( T \right)_\partial$ involves derivatives with respect to $x^-, x^i$ and hence does not act on $\partial_{\vec n}\partial_-^m\mathcal{O}_I(0)$.
On the other hand on the right hand side we find
\begin{align}
\bigg[T,  \sum_{I,{\vec n},m} C^{K, {\vec n},m}_{IJ}(x)(x)\partial_{\vec n}\partial_-^m\mathcal{O}_K(0)\bigg] & = 	\sum_{I,{\vec n},m}C^{K, {\vec n},m}_{IJ}(x)[T,\partial_{\vec n}\partial_-^m\mathcal{O}_K(0)]\nonumber\\
& =  	-\sum_{K,{\vec n},m}(\Delta_I+|\vec n|+2m) C^{K, {\vec n}}_{|{\vec n}|,m}(x)\partial_{\vec n}\partial_-^m\mathcal{O}_K(0)\ .
\end{align}
Comparing the two sides leads to the infinite series of differential equations satisfied by each of the coefficients $C^{K ,{\vec n},m}_{IJ}(x)$:
\begin{align}\label{scalingeq}
	 \left( T \right)_\partial C^{K ,{\vec n},m}_{IJ}(x) - (\Delta_K+|\vec n|+2m-\Delta_I-\Delta_J)C^{K ,{\vec n},m}_{IJ}(x) =0\ ,
\end{align}
which simply tell us how the coefficients $C^{K {\vec n},m}_{IJ}(x)$ need to scale as functions of $x^-, x^i$.

Let us now turn evaluate the commutator with $M_{i+}$. Since $[M_{i+},\mathcal{O}_J(0)]=0$ the left hand side is
\begin{align}
	[M_{i+},\mathcal{O}_I(x)\mathcal{O}_J(0) ]	 & = [M_{i+},\mathcal{O}_I(x)]\mathcal{O}_J(0)  \nonumber\\
& = -(\left( M_{i+} \right)_\partial \mathcal{O}_I(x) +\left(\tfrac{1}{2}\Delta_I \Omega_{ij} x^j - ip_I x^i	\right) \mathcal{O}_I(x))\mathcal{O}_J(0)  \nonumber\\
& =  -\sum_{K,{\vec n},m} \left( M_{i+}  \right)_\partial C^{K, {\vec n},m}_{IJ}(x)(x)\partial_{\vec n}\partial_-^m\mathcal{O}_K(0)\nonumber\\&\qquad\qquad- \left(\tfrac{1}{2}\Delta_I \Omega_{ij} x^j - ip_I x^i	\right)\sum_{K,{\vec n},m} C^{K, {\vec n}}_{|{\vec n}|,m}(x)\partial_{\vec n}\partial_-^m\mathcal{O}_K(0)\ . 
\end{align}
On the other hand
\begin{align}
\bigg[M_{i+},  \sum_{K,{\vec n},m} C^{K, {\vec n},m}_{IJ} (x)\partial_{\vec n}\partial_-^m\mathcal{O}_K(0)\bigg] & = 	\sum_{K,{\vec n},m} C^{K ,{\vec n},m}_{IJ}(x) [M_{i+},\partial_{\vec n}\partial_-^m\mathcal{O}_K(0)]\ .
\end{align}
To evaluate this we can first consider
\begin{align}
[M_{i+},\partial_{\vec n}\partial_-^m\mathcal{O}_K(x)]  & = \partial_{\vec n}\partial_-^m [M_{i+}, \mathcal{O}_K(x)]\nonumber\\
& = - \partial_{\vec n}\partial_-^m\left(\left( M_{i+} \right)_\partial \mathcal{O}_K(x) + \left(\tfrac{1}{2}\Delta_K \Omega_{ij} x^j - ip_K x^i	  \right) \mathcal{O}_K(x)\right)\ ,
\end{align}
and then set $x=0$. One can evaluate the first two terms to find
\begin{align}
	[M_{i+},\mathcal{O}_K(0)] & =0\nonumber\\
	[M_{i+},\partial_k\mathcal{O}_K(0)]& =-(\frac12 \Delta_K\Omega_{ik} - ip_K\delta_{ik})\mathcal{O}_K(0)
\end{align} 
Comparing the coefficients of  $\mathcal{O}_K(0)$ and $\partial_k\mathcal{O}_K(0)$ we can read off the equations
\begin{align}\label{Mi1}
 	\left(M_{i+} \right)_\partial C^{K,\vec 0, 0}_{IJ} + \left(\frac12\Delta_I\Omega_{ij}x^j- ip_Ix^i\right) C^{K, \vec 0,0}_{IJ} = -i\left(p_K\delta_{ij}+\frac{i}{2}\Delta_K\Omega_{ij}\right)C^{K,j,0}_{IJ}\ .
 	\end{align}
 We could keep going by including higher order descendants and their coefficients but this becomes increasingly tedious. It is clear that in general these equations can be used to determine $C^{K,\vec n,m}_{IJ}$ 
in terms of  $C^{K,\vec 0,0}_{IJ}$.

\subsection{A Special Case} \label{special}

Let us consider special case where
\begin{align} \label{constraint}
p_K = \pm \Delta_K/2R	\ .
\end{align}
Then we have 
\begin{align}
	\left(p_K\delta_{ij}-i\frac12\Delta_K\Omega_{ij}\right)\left(p_K\delta_{jk}+i\frac12\Delta_K\Omega_{jk}\right)=0\ ,
\end{align}
and hence from (\ref{Mi1}) we find the equation
\begin{align}\label{Smally}
\left(\delta_{ij}\mp i\frac12\Omega_{ij}\right)\left(	\left(M_{j+} \right)_\partial C^{K,\vec 0,0}_{IJ} + \frac12\Delta_I\Omega_{jk}x^kC^{K,\vec 0,0}_{IJ} - ip_Ix^j C^{K,\vec 0,0}_{IJ} \right)
=0\ .
\end{align}
The reason for this reduction is that for $p^2_K=\Delta_K^2/4R^2$
\begin{align}
\tilde {\cal O}_{iK}=p_K\partial_i{\cal O}_K +\frac{i}{2}\Delta_K \Omega_{ij}\partial_j{\cal O}_K	\ ,
\end{align}
is  also a primary operator if ${\cal O}_K$ is a scalar primary.  Said another way, an $\mathcal{O}_K$ with (\ref{constraint}) is the highest weight state of a special, short conformal multiplet.

Thus we find  a single first order differential equation for $C^{K,\vec 0,0}_{IJ}$ alone. 
To solve this we can assume $C^{K,\vec 0,0}_{IJ}$ is of the form $C^{K,\vec 0,0}_{IJ}=C^{K,\vec 0,0}_{IJ}(x^-,|x|^2)$.
  Indeed, this follows from the consistency of the OPE with the rotational symmetries.
If we define
\begin{align} \label{zdef}
z = x^- +\frac {i}{4R}|x|^2	\ ,
\end{align}
then the differential equation for $C^{K,\vec 0,0}_{IJ}$ reduces to
\begin{align} \label{specialsol}
z\partial C^{K,\vec 0,0}_{IJ} &=-(\Delta_I/2-p_IR)C^{K,\vec 0,0}_{IJ} \qquad +\ \text{sign}\nonumber \\
\bar z\bar \partial C^{K,\vec 0,0}_{IJ} &=-(\Delta_I/2+p_IR)C^{K,\vec 0,0}_{IJ}\qquad -\ \text{sign}
\ .
\end{align}
These are solved by
\begin{align}
C^{K,\vec 0,0}_{IJ} &= z^{-\tfrac12\Delta_I+ Rp_I}\tilde C^{K,\vec 0,0}_{IJ}(\bar z)\qquad +\ \text{sign}	\nonumber\\
C^{K,\vec 0,0}_{IJ} &= \bar z^{-\tfrac12\Delta_I- Rp_I}\tilde C^{K,\vec 0,0}_{IJ}( z)\qquad -\ \text{sign}
\end{align}
for any (anti) holomorphic function $\tilde  C^{K,\vec 0,0}_{IJ}$. However   we note that $C^{K,\vec 0,0}_{IJ} $ must satisfy (\ref{scalingeq}) which, in terms of $z,\bar z$ is
\begin{align}
2(z\partial + \bar z \bar\partial ) C^{K,\vec 0,0}_{IJ} =  ({\Delta_K-\Delta_I-\Delta_J}) C^{K,\vec 0,0}_{IJ}\ ,\end{align}
and this fixes $\tilde  C^{K,\vec 0,0}_{IJ}$ up to a multiplicative constant $c^K_{IJ}$. In particular  we find
\begin{align}\label{simpleC}
 C^{K,\vec 0,0}_{IJ} = c^K_{IJ}\left(\frac{1}{z\bar z}\right)^{\frac12\alpha_{IJ,K}}	\left(\frac{z}{\bar z}\right)^{ Rp_I \mp\tfrac12{\alpha_{IK,J}}}	\ ,
\end{align}
 where we introduced the notation 
 \begin{align}
 \alpha_{IJ,K} &= \tfrac12 \Delta_I+\tfrac12 \Delta_J-\tfrac12 \Delta_K\ .
 \end{align}
Hence, when a primary operator appearing in an OPE satisfies the constraint in \eqref{constraint}, we can determine the functional form of its OPE coefficient. In \cite{Golkar:2014mwa}, a similar argument was used to deduce the OPE coefficients of primary operators satisfying the unitarity bound $\Delta=d/2$ in theories with Schr\"odinger symmetry, where $d$ is the number of spatial dimensions. While the method we used is basically the same, the physical interpretation of our result is very different. First of all, the constraint in \eqref{constraint} leads to a first-order differential equation for the OPE coefficients, whereas  the unitarity bound leads to a second order differential equation. Noting that our correlators reduce to Schr\"odinger correlators with $d=4$ when $\Omega \rightarrow 0$, it would be interesting to investigate how the $\Omega$-deformed version of the unitarity bound constrains OPE coefficients, which is a different constraint than the one in \eqref{constraint}. 
 
 \section{Correlators from OPEs} \label{3pt}

Let us see how we can use the OPE coefficients to learn about correlation functions. First we will demonstrate the that the solution in \eqref{simpleC} can be used to derive the two-point scalar correlators reviewed in section \ref{review}. Next we will show that any three-point scalar correlator can be determined in terms of a certain OPE coefficient in the OPE of two operators in the correlator. If one of the operators in the correlator satisfies the constraint in \eqref{constraint}, it is then possible to determine the functional form of the three-point function and we will show that it agrees with the result of dimensionally reducing 6D Lorentzian correlators. Along the way, we will also derive an analogue of crossing symmetry for three-point correlators. 

We start be deriving scalar two-point functions from the OPE coefficients computed in \eqref{simpleC} by simply noting that
\begin{align}
\langle \mathcal{O}_I(x_1)\mathcal{O}_J(x_2)\rangle	& = \sum_{K,{\vec n},m} C_{IJ   }^{K,{\vec n}, m}(x_1,x_2)\langle\partial_{\vec n}\partial_-^m\mathcal{O}_K(x_2)\rangle\nonumber\\
& =  \sum_{{\vec n},m} C_{IJ   }^{0,{\vec n}, m}(x_1,x_2)\partial_{\vec n}\partial_-^m\langle\mathcal{O}_0(x_2)\rangle\nonumber\\ \ 
& =   C_{IJ   }^{0,{\vec 0}, 0}(x_1,x_2) \ .
\end{align}
Here we have used the fact that only the identity operator, denoted by $\mathcal{O}_0$, has a non-vanishing one-point function; $\langle\mathcal{O}_0(0)\rangle=1$. Furthermore the identity operator satisfies $p_0= \pm\Delta_0/2R=0$ so we can impose both signs in (\ref{Smally}) (indeed the left-hand-side vanishes without the projector) and hence we know  from (\ref{simpleC}) that 
\begin{align}
	 C_{IJ   }^{0,{\vec 0}, 0}(x_1,x_2) &= c^0_{IJ}\left(\frac{1}{z_{12}\bar z_{12}}\right)^{\frac12\alpha_{IJ,0}}	\left(\frac{z_{12}}{\bar z_{12}}\right)^{ Rp_I \mp\tfrac12{\alpha_{I0,J}}}\nonumber\\
	 & = c^0_{IJ}\left(\frac{1}{z_{12}\bar z_{12}}\right)^{\frac14(\Delta_I+\Delta_J)}	\left(\frac{z_{12}}{\bar z_{12}}\right)^{ Rp_I \mp\tfrac14(\Delta_I-\Delta_J) }\ .
\end{align}
Lastly we note that since  either choice of sign must work this requires   $\Delta_I=\Delta_J$, as is well-known for two-point functions. In this way we recover (\ref{2pts}), which was previously derived by solving the conformal Ward identities \cite{Lambert:2020zdc}.

Now let us consider three-point correlators. Taking the OPE of the first two operators in a three-point correlator gives
\begin{align} \label{3ptope}
  \langle \mathcal{O}_I(x_1)\mathcal{O}_J(x_2)\mathcal{O}_K(x_3)\rangle  & =   \sum_{L,\vec n,m}C^{L,\vec n,m}_{IJ}(z_{12},\bar z_{12})  \partial_{\vec n}\partial_-^m\langle \mathcal{O}_L(x_2)\mathcal{O}_K(x_3)\rangle\nonumber\\
& =  \sum_{I,\vec n,m} c_{KL}\delta_{\Delta_K,\Delta_L}\delta_{-p_K,p_L}C^{L,\vec n,m}_{IJ}(z_{12},\bar z_{12}) \nonumber\\ &\hskip2cm \times \partial_{\vec n}\partial_-^m\left[\left(\frac{1}{z_{23}\bar z_{23}}\right)^{\frac{\Delta_K}{2}}	\left(\frac{z_{23}}{\bar z_{23}}\right)^{-Rp_K}\right]\ ,
\end{align}
where the derivatives are with respect to $x_2$.
Thus, formally, there is a differential operator that acts on two-point functions to produce three-point functions. We have argued above that in principle all the $C^{K,\vec n,m}_{IJ}$ can be determined from $C^{K,\vec 0,0}_{IJ}$. On the other hand, the general solution to the conformal Ward identities in \eqref{3gen2} contains an unfixed function $H_{IJK}$. Thus the form of the three-point functions are fixed by symmetries of the theory and the unknown function $H_{IJK}$ is determined by the leading OPE coefficient $C^{K,\vec 0,0}_{IJ}$. 
More generally the operator appearing in the last line of \eqref{3ptope} acts on $n$-point functions to produce $(n+1)$-point functions and we recover the result, familiar from Lorentzian CFTs, that all the $n$-point functions are in principle determined by the OPE coefficients $C^{K,\vec n,m}_{IJ}$. Of course this is a formidable task in general. 

Recall that two and three-point functions in Lorentzian CFTs are completely fixed by conformal symmetry while four-point functions are determined up to an unknown function of conformal cross ratios. In this sense, we see that three-point functions of $SU(1,3)\times U(1)$ theories are analogous to four-point functions in Lorentzian CFTs. In Lorentzian CFTs, four-point functions are further constrained by crossing symmetry, which forms the foundation of the conformal bootstrap \cite{Ferrara:1973yt,Ferrara:1972kab,Polyakov:1974gs,Rattazzi:2008pe}. We may therefore expect a similar constraint to play a role for three-point functions of $SU(1,3)\times U(1)$ theories. To see how this works, we first take the limit $x_1\to x_2$ in \eqref{3ptope} keeping $z_{12}/\bar z_{12}$ and $z_{23}$ finite and arbitrary. If we write
\begin{align}
C^{K,\vec n, m}_{JK}(z_{12},\bar z_{12})  = x_{12}^{\vec n}(x_{12}^-)^m \bar{C}^{K,\vec n, m}_{JK} (z_{12},\bar z_{12}) \ ,
\end{align}
 where $x^{\vec n}$ is shorthand for a string of coordinates $x^ix^j \ldots$ of length $|\vec n|$ (it could also involve contributions from $\Omega_{jk}x^k$ in place of $x^j$) then
it  follows from (\ref{scalingeq}) that
\begin{align}
\bar{C}^{K,\vec n, m}_{JK}(z_{12},\bar z_{12})  =  \left( \frac{1}{z_{12}\bar z_{12}}\right)^{\tfrac12\alpha_{IJ,K}}F^{K,\vec n, m}_{JK}(z_{12}/\bar z_{12})	\ ,
\end{align}
for an unknown function $F^{K,\vec n, m}_{JK}(z_{12}/\bar z_{12})$.
Thus  the most singular term comes from $C^{K,\vec 0,0}_{IJ}$ and we get
\begin{align} \label{lim1}
 \lim_{x_1\to x_2} \langle \mathcal{O}_I(x_1)\mathcal{O}_J(x_2)\mathcal{O}_K(x_3)\rangle  & =   \sum_{L}C^{L,\vec 0, 0 }_{IJ}(z_{12},\bar z_{12})  \langle\mathcal{O}_L(x_2)\mathcal{O}_K(x_3)\rangle+\ldots\nonumber\\
& =  \sum_{L } c_{KL}\delta_{\Delta_L,\Delta_K}\delta_{-p_L,p_K}C^{L,\vec 0, 0}_{IJ}(z_{12},\bar z_{12})   \left(\frac{1}{z_{23}\bar z_{23}}\right)^{\frac{\Delta_K}{2}}	\left(\frac{z_{23}}{\bar z_{23}}\right)^{-Rp_K} +\ldots\nonumber\\
&=  \sum_{L } c_{KL}\delta_{\Delta_K,\Delta_L}\delta_{-p_L,p_K}F^{L,\vec 0 , 0 }_{IJ}\left(\frac{z_{12}}{\bar z_{12}}\right)  \left(\frac{1}{z_{12}\bar z_{12}}\right)^{\frac{\Delta_I+\Delta_J- \Delta_L}{4}}\nonumber\\ & \hskip5cm \times \left(\frac{1}{z_{23}\bar z_{23}}\right)^{\frac{\Delta_K}{2}}	\left(\frac{z_{23}}{\bar z_{23}}\right)^{-Rp_K} +\ldots\ ,
\end{align}
where the ellipsis are less singular terms.
 
On the other hand taking the same limit of (\ref{3gen2})   and noting that $z_{31}=-\bar z_{13}$ we find
\begin{align} \label{lim2}
\lim_{x_1\to x_2}\langle \mathcal{O}_I(x_1)\mathcal{O}_J(x_2)\mathcal{O}_K(x_3)\rangle 
&= \lim_{x_1\to x_2}	\delta_{-p_K,p_I+p_J}\left(\frac{1}{z_{12}\bar z_{12}}\right)^{\frac{\Delta_I+\Delta_J- \Delta_K}{4}}\left(\frac{1}{z_{23}\bar z_{23}}\right)^{\frac{\Delta_K}{2}}		\nonumber\\ &\qquad \times
\left(\frac{z_{12}}{\bar z_{12}}\right)^{\tfrac{R}{3}(p_I-p_J)}\left(\frac{z_{23}}{\bar z_{23}}\right)^{\frac{R}{3}(p_I+p_J-2p_K)} H_{IJK}\left(\frac{ z_{12} }{ \bar z_{12} }\right)\ .
\end{align}
The first term is diverging but it matches in both expressions. So do the second and fourth terms (since $p_I+p_J+p_K=0$). The $p$-conserving delta-functions also match in both expressions as   $p_L = p_I+p_J$.
Thus comparing \eqref{lim1} with \eqref{lim2} we read off 
\begin{align}\label{H3F}
 H_{IJK}\left(  \frac{z_{12}}{\bar z_{12}}\right)
&= \sum_L	c_{KL}\delta_{\Delta_K,\Delta_L}  \delta_{- p_L,p_K} F^{L,\vec 0 , 0}_{IJ}\left(\frac{z_{12}}{\bar z_{12}}\right)\left(\frac{z_{12}}{\bar z_{12}}\right)^{-\tfrac{R}{3}(p_I-p_J)}\ .\end{align}

Alternatively we can consider a different limit $x_2\to x_3$:
\begin{align}
 \lim_{x_2\to x_3} \langle \mathcal{O}_I(x_1)\mathcal{O}_J(x_2)\mathcal{O}_K(x_3)\rangle  & =   \sum_{L}C^{L,\vec 0,0 }_{JK}(z_{23},\bar z_{23})  \langle\mathcal{O}_I(x_1)\mathcal{O}_L(x_3)\rangle+\ldots\nonumber\\
& =  \sum_{L } c_{IL}\delta_{\Delta_I,\Delta_I}\delta_{-p_I,p^I}C^{L,\vec 0, 0}_{JK}(z_{23},\bar z_{23})   \left(\frac{1}{z_{13}\bar z_{13}}\right)^{\frac{\Delta_I}{2}}	\left(\frac{z_{13}}{\bar z_{13}}\right)^{Rp_I} +\ldots\nonumber\\
&=  \sum_{L } c_{IL}\delta_{\Delta_I,\Delta_L}\delta_{-p_I,p_L}F^{L,\vec 0 , 0}_{JK}\left(\frac{z_{23}}{\bar z_{23}}\right)  \left(\frac{1}{z_{23}\bar z_{23}}\right)^{\frac{\Delta_J+\Delta_K- \Delta_L}{4}}\nonumber\\ & \hskip5cm \times \left(\frac{1}{z_{13}\bar z_{13}}\right)^{\frac{\Delta_I}{2}}	\left(\frac{z_{13}}{\bar z_{13}}\right)^{Rp_I} +\ldots\ .
\end{align}
Again  taking the same limit of (\ref{3gen2})   we find
\begin{align}
\lim_{x_2\to x_3}\langle \mathcal{O}_I(x_1)\mathcal{O}_J(x_2)\mathcal{O}_3(x_3)\rangle 
&= \lim_{x_2\to x_3} \left(\frac{1}{z_{12}\bar z_{12}}\right)^{\frac{\Delta_I+\Delta_J-\Delta_K}{4}}	\left(\frac{1}{z_{23}\bar z_{23}}\right)^{\frac{-\Delta_I+\Delta_J+\Delta_K}{4}}	\left(\frac{1}{z_{13}\bar z_{13}}\right)^{\frac{\Delta_I-\Delta_J+\Delta_K}{4}}\nonumber\\ &\qquad \times
\left(\frac{z_{12}}{\bar z_{12}}\right)^{\tfrac{R}{3}(p_I-p_J)}\left(\frac{z_{23}}{\bar z_{23}}\right)^{\tfrac{R}{3}(p_J-p_K)}\left(\frac{z_{13}}{\bar z_{13}}\right)^{\tfrac{R}{3}(p_I-p_K)}	\nonumber\\
& \qquad\times H_{IJK}\left(\frac{z_{12}z_{23}z_{31}}{\bar z_{12}\bar z_{23}\bar z_{31}}\right)\delta_{-p_K,p_I+p_J}\nonumber\\
& = \lim_{x_2\to x_3}	\left(\frac{1}{z_{23}\bar z_{23}}\right)^{\frac{\Delta_J+\Delta_K-\Delta_I}{4}}	 \left(\frac{1}{z_{13}\bar z_{13}}\right)^{\frac{\Delta_I}{2}} 
 \left(\frac{z_{13}}{\bar z_{13}}\right)^{Rp_I}	\nonumber\\
& \qquad\times \left(\frac{z_{23}}{\bar z_{23}}\right)^{\tfrac{R}{3}(p_J-p_K)}H_{IJK}\left(\frac{ z_{23} }{ \bar z_{23} }\right)\delta_{-p_K,p_I+p_J}\ .
\end{align}
This time matching gives us
\begin{align} \label{crossed}
 H_{IJK}\left(  \frac{z_{23}}{\bar z_{23}}\right)
&= \sum_L	c_{IL}\delta_{\Delta_I,\Delta_L}  \delta_{ p_I,-p_L} F^{L,\vec 0 , 0}_{JK}\left(\frac{z_{23}}{\bar z_{23}}\right)\left(\frac{z_{23}}{\bar z_{23}}\right)^{-\tfrac{R}{3}(p_J-p_K)}\ .\end{align}
  However we must find the same function $H_{IJK}$ in both limits and  so we require that in general 
  \begin{align}\label{crossing}
 \sum_L	c_{KL}\delta_{\Delta_L,\Delta_K}  \delta_{- p_L,p_K} F^{L,\vec 0 , 0}_{IJ}\left(\zeta \right) \zeta^{-\tfrac{R}{3}(p_I-p_J)}=
  \sum_L	c_{IL}\delta_{\Delta_L,\Delta_I}  \delta_{- p_I,p_L} F^{L,\vec 0 , 0}_{JK}\left(\zeta\right) \zeta^{-\tfrac{R}{3}(p_J-p_K)}\ ,\end{align}
where $\zeta$ is the $SU(1,3)$-invariant cross-ratio defined below \eqref{3gen2}. We can think of this as a form of crossing symmetry for scalar three-point functions of $SU(1,3)\times U(1)$ theories. We expect that this relation holds for any consistent set of OPE coefficients. Indeed we will be able to check this in the special case below where both the first and third operators satisfy (\ref{constraint}) and hence the OPE coefficients are determined by (\ref{simpleC}).  Conversely in the general case it would be interesting to know if this constraint can be used to restrict or even determine the OPE coefficients. We leave this for future work.
  
 \subsection{Comparison with Dimensional Reduction}\label{subsec: dim red}
  
 Let us compare the above results with what we find from dimensional reduction. After Fourier expanding three-point scalar correlators of a 6D Lorentzian CFT along a compact null direction and comparing to \eqref{3gen2}, one finds that the function$ H_{IJK}$ takes the following form \cite{Lambert:2020zdc}:
 \begin{align} \label{Hreduced}
 H_{IJK}(\zeta) &= h\sum_{m=0}^\infty {-p_KR-\tfrac{\Delta_K}{2}+\alpha_{JK,I} - m -1\choose -p_KR-\tfrac{\Delta_K}{2}-m}{p_IR-\tfrac{\Delta_I}{2}+\alpha_{IJ,K}-m-1 \choose p_IR-\tfrac{\Delta_I}{2}-m}{\alpha_{KI,J}+m-1 \choose m} \nonumber\\
  &\hspace{20mm}\times \zeta^{-\tfrac{1}{2}\alpha_{KI,J}+\tfrac{1}{3}(p_IR-p_KR)-m}\ ,
\end{align}
where $h$ is a constant and assume that all $\Delta_I$ are even integers. Note that the formally infinite sum is in fact finite as the binomial coefficients vanish when the lower entry is negative.
 
One case we can  check is $p_K=-\Delta_K/2R$. In this case $H_{IJK}(\zeta)$ simplifies to 
\begin{align} \label{H3simple}
  &\hspace{-5mm}H_{IJK} (\zeta)
 = h\zeta^{-\tfrac{1}{2}\alpha_{KI,J} +\frac{R}{3}(p_I-p_K)}\ .
\end{align}
Since all the operators that appear in (\ref{H3F}) have $p_L=-p_K$ and $\Delta_L=\Delta_K$ we know that $p_L=\Delta_L/2R$ also. Therefore  $F^L_{IJ}$ must take the simple monomial form found in (\ref{simpleC}):
\begin{align}
F^{L,\vec 0 , 0}_{IJ}(\zeta) = f^{L }_{IJ} \zeta^{-\tfrac{1}{2}\alpha_{KI,J} +p_IR}	\ .
\end{align}
Thus we find, from (\ref{H3F}),
\begin{align}
	H_{IJK}(\zeta)&= \sum_L	c_{IL}\delta_{\Delta_L,\Delta_K}  \delta_{- p_L,p_K} F^{K,\vec 0 , 0}_{IJ}\left(\zeta\right)\zeta^{-\tfrac{R}{3}(p_I-p_J)} \nonumber\\
	&\propto 	\zeta^{-\tfrac{1}{2}\alpha_{KI,J} +\frac{R}{3}(p_I-p_K)}\ ,
\end{align}
in agreement with (\ref{H3simple}).

As another case   we can take $p_I = \Delta_I/2R$ where $H_{IJK}$ in (\ref{Hreduced}) also takes the form (\ref{H3simple}). Now we use the crossed form (\ref{crossed})
\begin{align}
	H_{IJK}(\zeta)&= \sum_L	c_{IL}\delta_{\Delta_L,\Delta_K}  \delta_{- p_L,p_I} F^{L,\vec 0 , 0}_{JK}\left(\zeta\right)\zeta^{-\tfrac{R}{3}(p_J-p_K)} \ .
\end{align}
We see that now $p_L=-p_I$ and $\Delta_L=\Delta_I$ so $p_L=-\Delta_L/2R$ and $ F^L_{JK}$ has just one term:
\begin{align}
 F^{L,\vec 0 , 0}_{JK} &= 	f^{L }_{JK} \zeta^{\tfrac{1}{2}\alpha_{IJ,K} +p_JR}\nonumber\\
 & = f^{L }_{JK} \zeta^{-\tfrac{1}{2}\alpha_{KI,J} +\frac12\Delta_I+p_JR}\nonumber\\
 & = f^{L }_{JK} \zeta^{-\tfrac{1}{2}\alpha_{KI,J} +p_IR+p_JR} \ ,
\end{align}
and hence
\begin{align}  
  \hspace{-5mm}H_{IJK} (\zeta) &\propto \zeta^{-\tfrac{1}{2}\alpha_{KI,J} +p_IR+p_JR -\frac{R}{3}(p_J-p_K)}\nonumber\\
   &= \zeta^{-\tfrac{1}{2}\alpha_{KI,J} +\frac{R}{3}(p_I-p_K)}\ .
\end{align}
 in agreement with (\ref{H3simple}). And hence it follows that if both $p_I=\Delta_I/2R$ and $p_K=-\Delta_K/2R$ then the crossing symmetry relation (\ref{crossing}) is satisfied (as one can also check directly). 

\section{Conclusions}\label{sect: Conclusions}
   
In this letter we have considered the operator product expansion for scalar operators in 5D field theories with an $SU(1,3)\times U(1)$ spacetime symmetry. In particular we showed that if a primary operator $\mathcal{O}_K$ appearing an OPE of primary operators $\mathcal{O}_I$ and $\mathcal{O}_J$ satisfies $p_K R = \pm\Delta_K/2$ (where $\Delta_K$ is its scaling dimension and $p_K$ is its $U(1)$ charge), then its OPE coefficient $C^{K,\vec 0, 0}_{IJ}$ can be determined in terms of a single constant. Furthermore, following the argument presented in \cite{Golkar:2014mwa} (see also \cite{Goldberger:2014hca}), we showed how the unknown function  $H_{IJK}(\zeta)$ that appears in the general solution to the three-point Ward identities can be determined from $C^{K,\vec 0, 0}_{IJ}$. In the special case $p_K R = \pm\Delta_K/2$ we were therefore able to determine $H_{IJK}(\zeta)$ and we found that it agrees with dimensional reduction of 6D Lorentzian conformal correlators. Thus at least one special class of three-point functions in 5D $SU(1,3)$ theories are necessarily those of a 6D  Lorentzian conformal field theory. This provides some hope that these 5D theories can be used to compute some quantities in more traditional, but non-Lagrangian, 6D theories such as the famous   $(2,0)$ theory. 

In future work we hope to extend our results to more general classes of operators. As a first step, it would be important to better understand the physical interpretation of the 5D OPE studied in this paper. Indeed, since local operators in the 5D theory correspond to Fourier modes along an internal null direction they are non-local operators in six dimensions. Moreover their Kaluza-Klein numbers in six dimensions correspond to instanton numbers in five dimensions. If we construct 5D operators by taking traces of products of fundamental scalar fields $X^I$ dressed with instanton operators, the constraint $p_K R = \pm\Delta_K/2$ corresponds to attaching a single instanton or anti-instanton to each scalar field (which has classical scaling dimension equal to two). In \cite{Lambert:2020zdc}, this was shown to be the minimal number of instantons needed to have nonzero two-point functions. More generally, we could relax this constraint by attaching more instantons to each scalar field, which would correspond to probing higher modes along the null direction. 
We may then explore how to generalise the solution for $C^{K,\vec 0, 0}_{IJ}$ when $p_K R = \pm\Delta_K/2$ to other cases, for example when an operator in the OPE satisfies an $\Omega$-deformed analogue of the unitarity bound $\Delta=d/2$, which was shown to fix OPE coefficients in theories with Schr\"odinger symmetry in \cite{Golkar:2014mwa}. It would also be interesting to explore whether the crossing symmetry relation for three-point functions found in \eqref{crossing} provides the starting point for a non-relativistic conformal bootstrap. We hope to address these exciting questions in the future.
 
\section*{Acknowledgements}
 
We'd like to thank Parijat Dey for helpful comments on a draft version of this paper. N.L. is supported in part  by an STFC consolidated grant ST/X000753/1.  A.L. is supported by an STFC Consolidated Grant ST/T000708/1. R.M. is supported by David Tong's Simons Investigator award.

\end{document}